\documentclass[aps,prb,showpacs,superscriptaddress,twocolumn]{revtex4}
\usepackage{amsmath}
\usepackage{amsfonts}
\usepackage{amssymb}
\usepackage{graphicx}%

\begin{document}
\title{Two-dimensional quantum rings with oscillating spin-orbit interaction
strength:\\ A wave function picture}
\author{P\'{e}ter F\"{o}ldi}
\email{foldi@physx.u-szeged.hu}
\affiliation{Department of Theoretical Physics, University of Szeged, Tisza Lajos
k\"{o}r\'{u}t 84, H-6720 Szeged, Hungary}
\author{Orsolya K\'{a}lm\'{a}n}
\affiliation{Department of Quantum Optics and Quantum Information, Research Institute for
Solid State Physics and Optics of the Hungarian Academy of Sciences,
Konkoly-Thege Mikl\'{o}s \'{u}t 29-33, H-1121 Budapest, Hungary}
\author{Mih\'{a}ly G. Benedict}
\affiliation{Department of Theoretical Physics, University of Szeged, Tisza Lajos
k\"{o}r\'{u}t 84, H-6720 Szeged, Hungary}

\begin{abstract}
We determine the relevant spinor valued wave functions for a two-dimensional
quantum ring in the presence of Rashba-type spin-orbit interaction (SOI). The case of
constant SOI strength is considered first, then we investigate the physical consequences of
time-dependent (oscillating) SOI strength. Floquet's method is applied to find time-dependent
eigenspinors, and it is shown that the Floquet quasienergies and thus their
differences (the generalized Rabi frequencies) are determined by the radial
boundary conditions. Time evolution of various initial states is calculated.

\end{abstract}

\maketitle

\section{Introduction}
Quantum rings made of semiconducting materials exhibiting Rashba-type\cite{R60} spin-orbit interaction\cite{G00,NATE97,SKY01,KTHS06} (SOI) have attracted considerable attention due to fundamental spin-dependent quantum interference phenomena that are observable in these systems. Since the strength of the SOI  can be tuned with external gate voltages,\cite{NATE97} quantum rings or systems of them\cite{BKSN06,ZW07,KFBP08b,KFBP08c,CSz08} also have possible spintronic\cite{ALS02} applications. An interesting point in this context is the question of time-dependent SOI strength.  Studies of transport related problems with oscillating SOI have been initiated in Ref.~[\onlinecite{WC07}] for a ring, and in Ref.~[\onlinecite{RCM08}] for a ring-dot system, mainly in the context of spin currents. An analytical one-dimensional model for this problem has been developed in Ref.~[\onlinecite{FBKP09}]. In the current paper we consider a ring with finite width, calculate the time-dependent eigenspinors, i.e, Floquet\cite{MB02,LR99} states of the problem with oscillating SOI strength, and investigate the dynamics of various initial states. Our method is based on the solution of analytic equations leading to the exact spinor valued wave functions and providing the maximal possible insight into the dynamical processes. This approach allows us to see clearly the appearance of high harmonics of the driving SOI oscillations in the Floquet states and consequently also in the position and spin-resolved time evolution of arbitrary initial states.

Considering different two-dimensional (2D) approaches, magnetotransport in a finite-width ring with Rashba SOI has been investigated using the scattering matrix method, by dividing the device into small stripes which were then connected by
scattering matrices.~\cite{WC06} In Ref.~[\onlinecite{NS09}] the single-electron spin orbitals of a
two-dimensional ring were found by diagonalization of the Hamiltonian in a basis of multicenter
Gaussian functions. The model of a two-dimensional hydrogen atom in the presence of Rashba SOI was
used in Ref.~[\onlinecite{G08}] to describe electronic bound states around charged impurities in
two-dimensional systems. Persistent currents in semiconductor ballistic rings with spin-orbit Rashba
interaction were also investigated by the multicomponent Tomonaga-Luttinger liquid
model.~\cite{PG04} Luttinger model taking heavy and light-hole states into account has recently been applied for the investigation of spin-related quantum phases.\cite{JZ10} For the theoretical description of transport properties of diametrically connected finite-width rings with Rashba SOI, a tight-binding model with concentric lattice of ring chains has been used\cite{SN04}, while in Ref. [\onlinecite{FR04}] a spin-dependent recursive Green-function technique
was applied to the relevant 2D Hamiltonian. Magnetic field related effects in ring shaped objects with finite width have also been investigated intensively.\cite{SzP09,SP09,PSz10,BO10a,BO10b,A04}
Our approach is similar to that of Ref.~[\onlinecite{FL09}], where the features
of certain low-lying states as a function of the (constant) Rashba SOI were investigated, and
Refs.~[\onlinecite{BS01,TLG04}], where a circular (not ring shaped) quantum dot was considered.

In Sec.~\ref{Bulksec} we recall the eigenstates of the Hamiltonian with Rashba-type SOI term in an infinite 2D space ("bulk 2D" eigenspinors) then it is shown how the radial boundary conditions lead to a discrete spectrum (Subec.~\ref{CSOIsubsec}). We transfer this method to the case of oscillating SOI strength in Subsec.~\ref{TDSOIsubsec} where we determine the Floquet quasienergy spectrum as well as the corresponding Floquet states that satisfy the radial boundary conditions. As applications, in Sec.~\ref{TDsec} spin and charge density oscillations and the dynamics of wave packets are calculated. Summary is given in Sec.~\ref{Sumsec}.

\section{Spin-dependent eigenvalue problem in 2D}
\label{Bulksec}
The Hamiltonian for an electron moving in the $x-y$ plane in the presence of
Rashba spin-orbit interaction, can be written as
\begin{align}
H  &  =\frac{\hbar^{2}}{2m^{\ast}}\left(  P_{r}^{2}+\frac{1}{r^{2}}L_{z}%
^{2}\right)  +2\alpha\left(  \frac{1}{r}S_{r}L_{z}-S_{\varphi}P_{r}\right)
\nonumber\\
&  =H_{0}+H_{\mathrm{SO}}, \label{H}%
\end{align}
using cylindrical coordinates. Here $m^{\ast}$ is the effective mass of the
electron, $\alpha$ is the Rashba coupling strength, $H_{0}$ is the
spin-independent part and $H_{\mathrm{SO}}$ contains the spin-orbit
interaction term. The radial and azimuthal spin operators, $S_{r}$ and $S_{\varphi}$, are
dimensionless here (their true dimension comes from with the factor $\hbar$).
We have also introduced the notation $P_{r}=-i\partial/\partial r$ and
$L_{z}=-i\partial/\partial\varphi$ for the radial component of the
momentum and $z$ component of the dimensionless orbital angular momentum, respectively.
Nanoscale quantum rings, for which the Hamiltonian above is relevant,
can be fabricated from e.g.~InAlAs/InGaAs based heterostructures\cite{KNAT02} or
HgTe/HgCdTe quantum wells.\cite{KTHS06} Let us note that the confining potential does not appear explicitly in the Hamiltonian above. We are going to consider hard wall boundary conditions that will be taken into account in Sec.~\ref{boundsec}. Although soft wall confinement is definitely more realistic (see E.g.~Ref.~[\onlinecite{Bu90}]), we expect that our analytic approach can capture the most important physical phenomena appearing in rings of finite width.

The Hamiltonian of Eq. (\ref{H}) commutes with
\begin{equation}
K=L_{z}+S_{z},
\end{equation}
the $z$ component of the total angular momentum as obviously $[H_{0},K]=0,$
while in
\begin{equation}
\left[  H_{\mathrm{SO}},K\right]  =2\alpha\left[  \frac{1}{r}S_{r}%
L_{z}-S_{\varphi}P_{r},L_{z}+S_{z}\right]  \label{HSOK}%
\end{equation}
we obtain for the nonzero commutators $\left[  S_{r},L_{z}\right]  =iS_{\varphi}%
$, $\left[  S_{r},S_{z}\right]  =
-iS_{\varphi}$, $\left[  S_{\varphi},L_{z}\right]  =-iS_{r}$ and $\left[
S_{\varphi},S_{z}\right]  =
iS_{r}$.
This can also be expected from the fact that $H$ comes from the Dirac
Hamiltonian that conserves total angular momentum. Note that $L_{z}$ and
$S_{z}$ alone are not constants of motion.

Looking at the form of $K,$ one sees that its eigenvalues $\kappa$ are doubly
degenerete, because an eigenvalue $m$ of $L_{z}$ and $1/2$ \ of $S_{z}$ leads to
$\kappa=m+1/2,$ which gives the same result as eigenvalues $m+1$ and $-1/2$ of
the respective operators. This is similar to the one-dimensional case, where
the eigenvectors of the Hamiltonian in the subspace corresponding to a given value of $\kappa$ are
linear combinations of the eigenstates $e^{im\varphi}\left\vert \uparrow\right\rangle$
and $e^{i(m+1)\varphi}\left\vert \downarrow\right\rangle$.\cite{BS01,TLG04,MPV} Note that in the eigenbasis of the spin oparator $S_z$ we have $\left\vert \uparrow\right\rangle= \begin{pmatrix} 1 \\ 0 \end{pmatrix}$ and $\left\vert \downarrow\right\rangle= \begin{pmatrix} 0 \\ 1 \end{pmatrix}.$

Guided by the cylindrical symmetry of the system, in the two-dimensional case
we are led to look for the solution of the eigenvalue equation of
$H_{0}+H_{\mathrm{SO}}$ in the subspace corresponding to a given value of
$\kappa$ in the form of a linear combination of basis states $\phi
_{m}\left\vert \uparrow\right\rangle $ and $\phi_{m+1}\left\vert \downarrow\right\rangle $
where $\phi_{m}=Z_{m}(kr)e^{im\varphi}$ are the eigenfunctions of the free
Hamiltonian $H_{0}=-\frac{\hbar^{2}}{2m^{\ast}}\Delta_{r\varphi}$
corresponding to an eigenvalue $\frac{\hbar^{2}}{2m^{\ast}}k^{2}$ and written
here in terms of cylindrical coordinates.\cite{BS01} As it is well known, the cylindrical
wave functions $Z_{m}(kr)$ obey the Bessel equation:
\begin{equation}
\left(  -\frac{\partial^{2}}{\partial r^{2}}-\frac{1}{r}\frac{\partial
}{\partial r}+\frac{m^{2}}{r^{2}}\right)  Z_{m}=k^{2}Z_{m},
\end{equation}
where $k$ is a positive real number, while in the case of a closed ring, $m$
must be an integer. Accordingly we look for the eigenfunctions of the
Hamiltonian as a linear combination $\psi=a_{\kappa}\phi_{m}\left\vert
\uparrow\right\rangle +b_{\kappa}\phi_{m+1}\left\vert \downarrow\right\rangle ,$ which leads
to the following equation for the eigenspinors:
\begin{align}
&  \left(  \!%
\begin{array}
[c]{cc}%
-\Delta_{r\varphi} & \gamma e^{-i\varphi}\left(  \frac{\partial}{\partial
r}-i\frac{1}{r}\frac{\partial}{\partial\varphi}\right) \\
\gamma e^{i\varphi}\left(  -\frac{\partial}{\partial r}-i\frac{1}{r}%
\frac{\partial}{\partial\varphi}\right)  & -\Delta_{r\varphi}%
\end{array}
\!\right)  \!%
\begin{pmatrix}
a_{\kappa}\phi_{m}\\
b_{\kappa}\phi_{m+1}%
\end{pmatrix}
\nonumber\\
&  =\frac{2m^{\ast}}{\hbar^{2}}E%
\begin{pmatrix}
a_{\kappa}\phi_{m}\\
b_{\kappa}\phi_{m+1}%
\end{pmatrix},
\label{HPSI}
\end{align}
where $\gamma=2m^{\ast}\alpha/\hbar^{2}$. Due to the recurrence relations
valid for Bessel functions \cite{GR71,AS65} one has
\begin{align}
e^{-i\varphi}  &  \left(  \frac{\partial}{\partial r}-i\frac{1}{r}%
\frac{\partial}{\partial\varphi}\right)  Z_{m+1}e^{i(m+1)\varphi}%
=kZ_{m}e^{im\varphi},\nonumber\\
e^{i\varphi}  &  \left(  -\frac{\partial}{\partial r}-i\frac{1}{r}%
\frac{\partial}{\partial\varphi}\right)  Z_{m}e^{im\varphi}=kZ_{m+1}%
e^{i(m+1)\varphi},
\end{align}
thus
\begin{align}
H_{\mathrm{SO}}\phi_{m}\left\vert \uparrow\right\rangle  &  =\gamma k\phi
_{m+1}\left\vert \downarrow\right\rangle ,\\
H_{\mathrm{SO}}\phi_{m+1}\left\vert \downarrow\right\rangle  &  =\gamma k\phi
_{m}\left\vert \uparrow\right\rangle ,
\end{align}
and Eq. (\ref{HPSI}) results in
\begin{equation}
\frac{\hbar^{2}}{2m^{\ast}}\left[
\begin{array}
[c]{cc}%
k^{2} & \gamma k\\
\gamma k & k^{2}%
\end{array}
\right]  \left[
\begin{array}
[c]{c}%
a_{\kappa}\\
b_{\kappa}%
\end{array}
\right]  =E\left[
\begin{array}
[c]{c}%
a_{\kappa}\\
b_{\kappa}%
\end{array}
\right]  ,
\end{equation}
where the $\left[.\right]$ brackets denote that we have written the eigenvalue equation in the
$\left\lbrace\phi_{m}\left|\uparrow\right>,\phi_{m+1}\left|\downarrow\right>\right\rbrace$ basis.
This yields the following two eigenvalues
\begin{equation}
E^{\pm}=\frac{\hbar^{2}}{2m^{\ast}}\left(  k^{2}\pm\gamma k\right)  ,
\label{energies}%
\end{equation}
and the corresponding ratios of the coefficients are:
\begin{equation}
\left(  \frac{a_{\kappa}}{b_{\kappa}}\right)  ^{\pm}=\pm1. \label{constapb}%
\end{equation}
The actual values of the coefficients can be determined by an additional
normalization procedure to be given below. The result above is valid for
arbitrary positive, zero or negative integer $m$.

\begin{figure}[tbh]
\includegraphics[width=5cm]{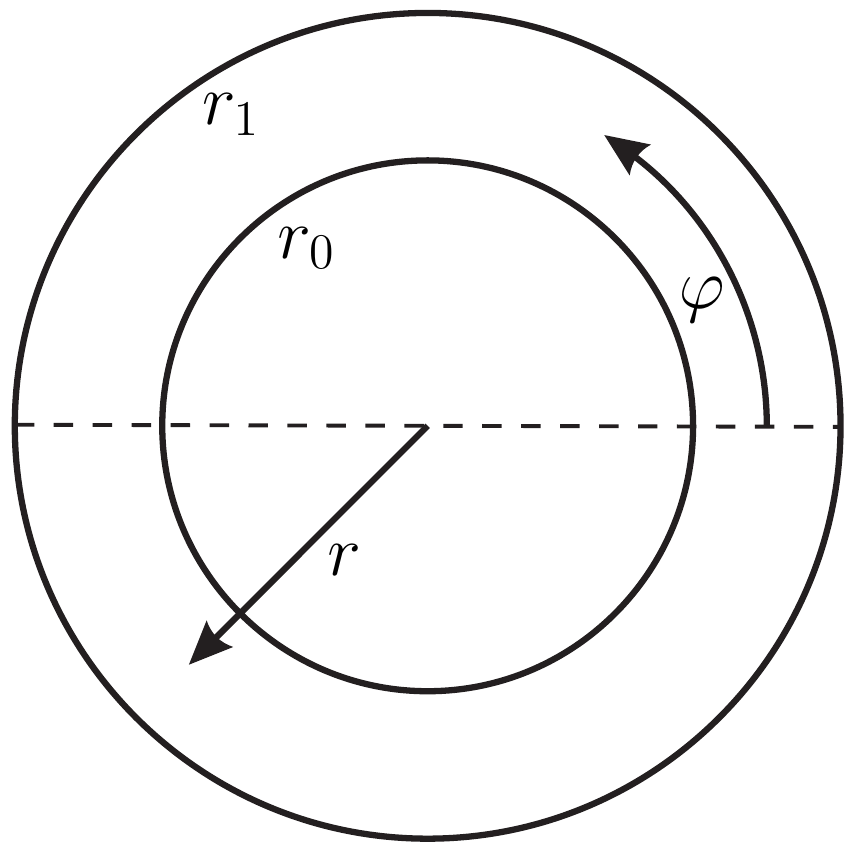}\caption{Two-dimensional quantum ring and the relavant coordinates.}%
\label{ringfig}%
\end{figure}

\section{Boundary conditions}
\label{boundsec}
The geometry shown in Fig. \ref{ringfig} requires to specify the solution
of the spin-dependent eigenvalue problem presented in the previous section
with appropriate boundary conditions. In the following we consider hard wall
potentials that confine the electrons into the ring shaped region. We assume that
the spinor valued wave functions $\psi$ vanish at the radial coordinates
$r_{0}$ and $r_{1}$:
\begin{equation}
\psi(kr_{0})=\psi(kr_{1})=%
\begin{pmatrix}
0\\
0
\end{pmatrix}
.\label{boundary}%
\end{equation}
From now on, we use dimensionless quantities. With the notation $\Omega
=\frac{\hbar}{2m^{\ast}r_{1}^{2}}$ energy is to be measured in units of
$\hbar\Omega,$ while the strength of the Rashba spin-orbit interaction is
characterized by the dimensionless ratio $\omega/\Omega=2m^{\ast}\alpha
r_{1}/\hbar^{2}=\gamma r_{1}$.

\subsection{Constant SOI}
\label{CSOIsubsec}
Results of Sec.~II show that a certain energy $\varepsilon$ can result from two positive $k$
values, which are given by the solutions of Eq.~(\ref{energies})
\begin{equation}
\begin{aligned}
&k_{+}=-\frac{\omega}{2\Omega r_{1}}+\frac{1}{r_{1}}\sqrt{\left(  \frac{\omega
}{2\Omega}\right)  ^{2}+\varepsilon}, \\
&k_{-}=\frac{\omega}{2\Omega r_{1}%
}+\frac{1}{r_{1}}\sqrt{\left(  \frac{\omega}{2\Omega}\right)  ^{2}%
+\varepsilon}.
\end{aligned}
\end{equation}
where $\varepsilon=E/\hbar\Omega.$ For a given value of $\kappa=m+1/2,$ we
obtain four linearly independent spinors \cite{BS01,FL09,TLG04}
\begin{equation}
\begin{aligned} &|\psi_1\rangle=\begin{pmatrix} J_m(k_+r) e^{im\varphi} \\ J_{m+1}(k_+r) e^{i(m+1)\varphi} \end{pmatrix}, \\ &|\psi_2\rangle=\begin{pmatrix} N_m(k_+r) e^{im\varphi} \\ N_{m+1}(k_+r) e^{i(m+1)\varphi} \end{pmatrix}, \\ &|\psi_3\rangle=\begin{pmatrix} -J_m(k_-r) e^{im\varphi} \\ J_{m+1}(k_-r) e^{i(m+1)\varphi} \end{pmatrix}, \\ &|\psi_4\rangle=\begin{pmatrix} -N_m(k_-r) e^{im\varphi} \\ N_{m+1}(k_-r) e^{i(m+1)\varphi} \end{pmatrix}, \label{const4} \end{aligned}
\end{equation}
the superpositions of which correspond to the energy $\varepsilon$. However,
when we take the boundary conditions into account, generally we find that they
cannot be satisfied for an arbitrary energy. Equations~(\ref{boundary}) for
$\Psi=\sum c_{n}\left\vert \psi_{n}\right\rangle $ mean a set of four linear
equations for the expansion coefficients $c_{n}$, and nontrivial solution
exists only if the determinant
\begin{equation}
\begin{aligned} &D\left(\varepsilon,\frac{\omega}{\Omega}\right)=\\ &\begin{vmatrix} J_m\left(k_+ r_0\right) & N_m\left(k_+ r_0\right) & -J_m\left(k_- r_0\right) & -N_m\left(k_- r_0\right)\\ J_{m+1}\left(k_+ r_0\right) & N_{m+1}\left(k_+ r_0\right) & J_{m+1}\left(k_- r_0\right) & N_{m+1}\left(k_- r_0\right)\\ J_m\left(k_+ r_1\right) & N_m\left(k_+ r_1\right) & -J_m\left(k_- r_1\right) & -N_m\left(k_- r_1\right) \\ J_{m+1}\left(k_+ r_1\right) & N_{m+1}\left(k_+ r_1\right) & J_{m+1}\left(k_- r_1\right) & N_{m+1}\left(k_- r_1\right) \end{vmatrix}\end{aligned}\label{constdet}%
\end{equation}
vanishes.
\begin{figure}[tbh]
\includegraphics[width=8cm]{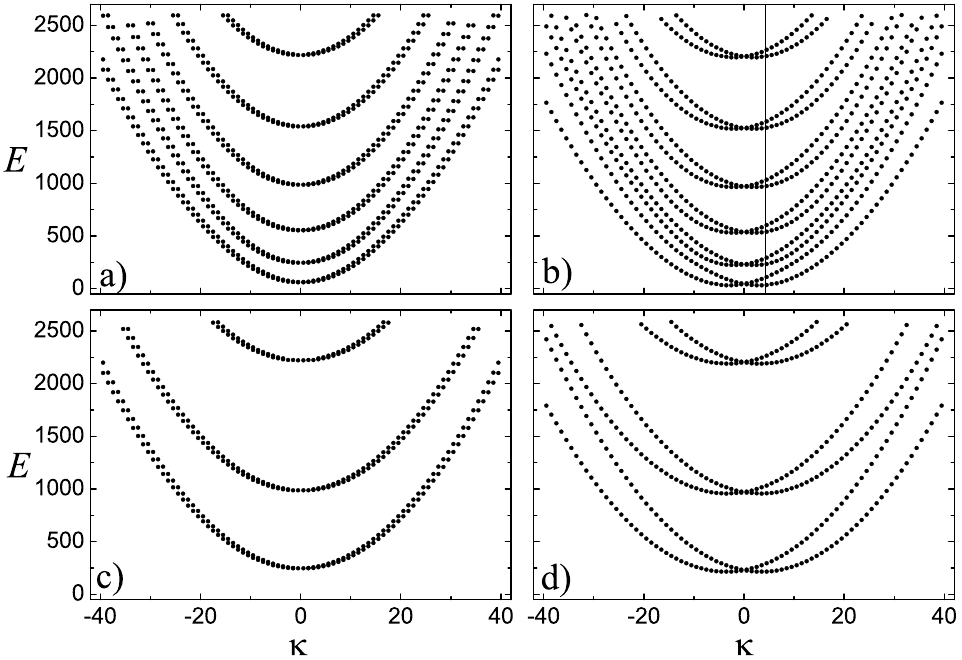}\caption{Level scheme for two-dimensional rings with different parameters. $r_0/r_1=0.6, \omega/\Omega=0.1$ (a); $r_0/r_1=0.6, \omega/\Omega=4.0$ (b); $r_0/r_1=0.8, \omega/\Omega=0.1$ (c) and $r_0/r_1=0.8, \omega/\Omega=4.0$ (d). Energy is measured in units of $\hbar\Omega=\frac{\hbar^2}{2m^{\ast}r_{1}^{2}}.$ The vertical line in panel (b) corresponds to $\kappa=9/2,$ which is the value for which the wave functions are plotted in Fig.~\ref{wffig}. $1/\hbar\Omega.$}
\label{spectrfig}%
\end{figure}
(Note that energy and SOI dependence of $D$ results from that of the
wave numbers $k_{\pm}$.) For a certain (constant) SOI strength, by sweeping the energy,
we can look for zeros of $D(\varepsilon),$ to find a discrete spectrum
determined by the boundary conditions, as demonstrated in Fig.~\ref{spectrfig}.
Different double "curves" (that are close to parabolas) represent different
radial modes: the ones with the lowest energy correspond to states with no
node between $r_{0}$ and $r_{1},$ while for states with energies $n$ step higher
there are $n$ circles between $r_{0}$ and $r_{1}$ where both spinor components
are zero (see Fig.~\ref{wffig}). The two states
corresponding to the same radial mode and value of $\kappa$ are orthogonal if we take the spin degree of freedom also into account. The eigenenergies
and states can be labeled as $\varepsilon_{n\kappa}^{\pm},$ $|\psi_{n\kappa
}^{\pm}\rangle,$ respecively, with $\kappa$ and $n$ referring to the spatial
degrees of freedom (azimuthal and radial coordinates) while $+$ and $-$
distinguish the two spin directions.

In this way we can obtain a complete solution of the eigenvalue problem
related to the Hamiltonian (\ref{H}) with the boundary conditions described by
Eq.~(\ref{boundary}) for any fixed SOI strength. This allows us to calculate
the time evolution of any initial state $|\Psi(0)\rangle=\sum_{n\kappa}%
\beta_{n\kappa}^{+}|\psi_{n\kappa}^{+}\rangle+\beta_{n\kappa}^{-}%
|\psi_{n\kappa}^{-}\rangle$ simply as
\begin{equation}
|\Psi(\tau)\rangle=\sum_{n\kappa}\beta_{n\kappa}^{+}|\psi_{n\kappa}^{+}\rangle
e^{-i\varepsilon_{n\kappa}^{+}\tau}+\beta_{n\kappa}^{-}|\psi_{n\kappa}%
^{-}\rangle e^{-i\varepsilon_{n\kappa}^{-}\tau}.
\end{equation}
We use here the dimensionless time variable $\tau=\Omega t/2\pi,$ and the
expansion coefficients are given by the inner product:
\begin{equation}
\begin{aligned}
\beta_{n\kappa}^{\pm}&=\langle\langle\psi_{n\kappa}^{\pm}|\Psi(0)\rangle\rangle \\
&=\int\limits_{r_{0}}^{r_{1}}\int\limits_{0}^{2\pi}  \langle \psi_{n\kappa}%
^{\pm}(r,\varphi))\vert \Psi(r,\varphi,0)\rangle  rd\varphi dr,
\label{inner}
\end{aligned}
\end{equation}
where $\langle .\vert .\rangle$ denotes the usual spin inner product.
Note that the eigenstates are normalized in the sense of Eq.~(\ref{inner}), i.e., $\langle\langle\psi_{n\kappa}^{+}|\psi_{n\kappa}^{+}\rangle \rangle
=\langle\langle\psi_{n\kappa}^{-}|\psi_{n\kappa}^{-}\rangle \rangle=1.$

The spinor valued eigenfunctions are visualized in Fig.~\ref{wffig}, where the spin independent probability density
\begin{equation}
\rho^{\pm}(r,\varphi)=
\langle\psi_{n\kappa}^{\pm}(r,\varphi) \vert \psi_{n\kappa}^{\pm}(r,\varphi)\rangle
\label{dens}
\end{equation}
is shown for $n=1,\ldots,4$ as well as the position-dependent expectation values $\langle S_x\rangle,$ $\langle S_y \rangle$ and $\langle S_z \rangle,$
where e.g.
\begin{equation}
\langle S_z \rangle(r,\varphi)=\langle\psi_{n\kappa}^{+}(r,\varphi) \vert S_z \vert \psi_{n\kappa}^{+}(r,\varphi)\rangle.\end{equation}

\begin{figure}[tbh]
\includegraphics[width=8cm]{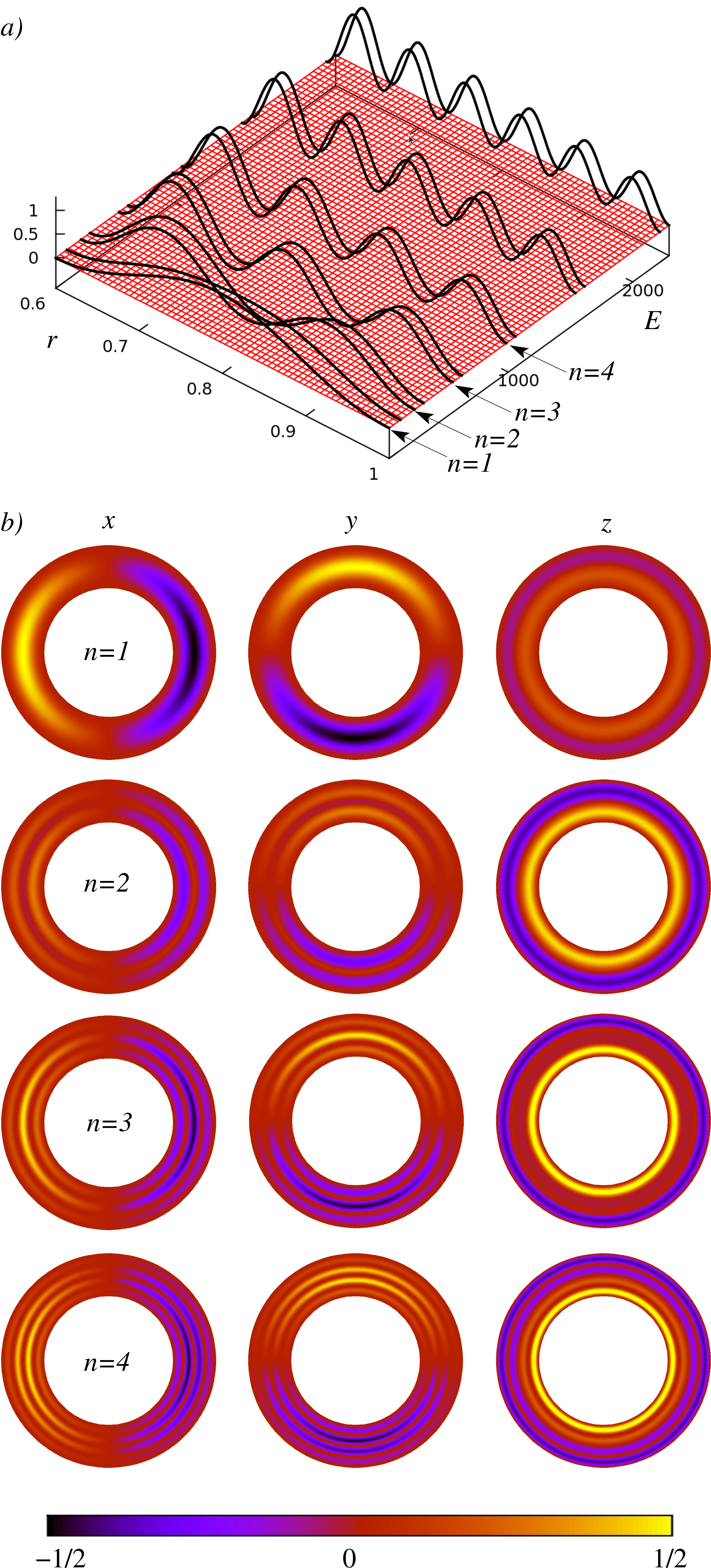}\caption{Wave functions for $\kappa=9/2, r_0/r_1=0.6, \omega/\Omega=4.0$ [See the vertical line in Fig.~\ref{spectrfig}(b)]. Probability densities (\ref{dens}) -- that do not depend on $\varphi$ in the current cases -- are shown in panel (a), while the position-dependent expectation value of $S_x, S_y$ and $S_z$ for the states indicated by the arrows can be seen in panel (b).}%
\label{wffig}%
\end{figure}

\subsection{Oscillating SOI strength}
\label{TDSOIsubsec}
When the strength of the SOI is not constant, the dimensionless Hamiltonian
still can be transformed into an algebraic matrix using the basis states $\phi
_{m}\left\vert \uparrow\right\rangle $ and $\phi_{m+1}\left\vert \downarrow\right\rangle $, leading to
\begin{equation}
H(\tau)=\left[
\begin{array}
[c]{cc}%
k^{2} & k\frac{\omega(\tau)}{\Omega}\\
k\frac{\omega(\tau)}{\Omega} & k^{2}\label{redtdHam}%
\end{array}
\right],
\end{equation}
where units of $1/r_1$ have been used for the wave number $k.$
In the following we consider the case of oscillating SOI strength, i.e.,
\begin{equation}
\frac{\omega(\tau)}{\Omega}=A\cos(\nu\tau)=A\cos\left(\frac{2\pi \tau}{\mathcal{T}}\right).
\end{equation}
According to Floquet's theorem, there is a time-dependent basis
\begin{align}
&  |\Phi_{r}(\tau)\rangle=|u_{r}(\tau)\rangle e^{-i\epsilon_{r}\tau},\ \ |u_{r}%
(\tau+\mathcal{T})\rangle=|u_{r}(\tau)\rangle,\nonumber\\
&  \langle u_{1}(\tau)|u_{2}(\tau)\rangle=0,\ \ \langle u_{r}(\tau)|u_{r}%
(\tau)\rangle=1,\label{Fbasis}%
\end{align}
where $\mathcal{T}=2\pi/\nu$. Unlike the Floquet states $|u_{r}%
(\tau)\rangle,$ the elements of this basis themselves are not $\mathcal{T}%
$-periodic functions. Let us recall that if $\epsilon_{r}$ is a Floquet quasi-energy
and the corresponding state is $|\Phi_{r}(\tau)\rangle,$ then the same holds for
$\epsilon_{r}+n\nu$ and $|\Phi_{r}(\tau)\rangle\times\exp(in\nu
\tau)$ for any integer $n.$ However, these states are equivalent from the
dynamical point of view, thus it is sufficient to focus on only two
nonequivalent quasi-energies.

\begin{figure}[tbh]
\includegraphics[width=8cm]{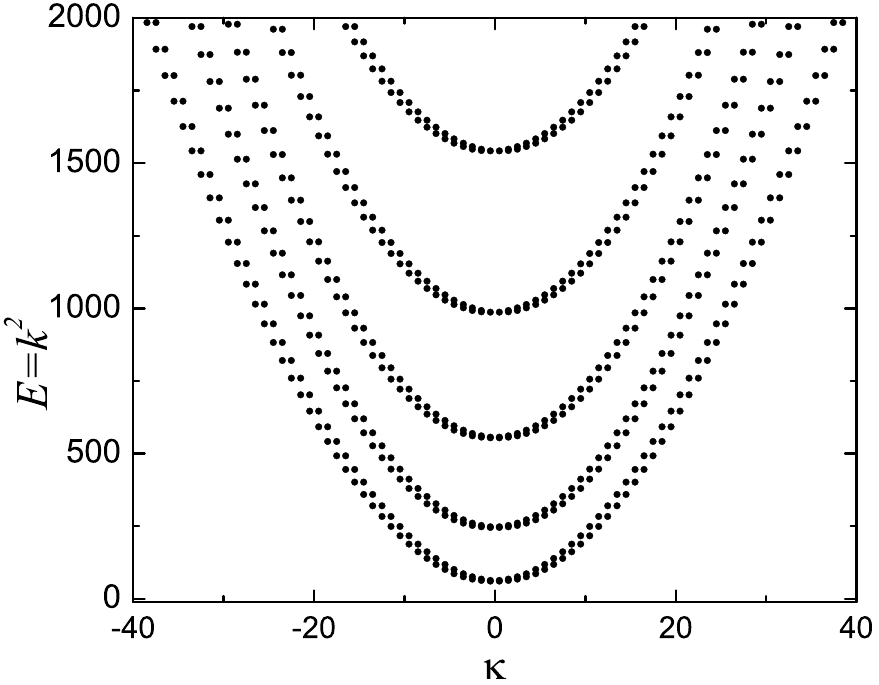}\caption{Floquet quasienergies $k^2$ (measured in units of $\hbar\Omega$) for a ring with $r_0/r_1=0.6$.}%
\label{tdspectrfig}%
\end{figure}

The determination of the two relevant elements of the basis (\ref{Fbasis}) and the
corresponding Floquet quasienergies for a given value of $k$ can be done
numerically essentially by the Fourier expansion of the eigenvalue equation.
In the current case, however,  an exact analytical result can also be found
(see Appendix A):
\begin{equation}
\begin{aligned} &|\Phi^+ (\tau)\rangle=e^{-i k^2 \tau}\left[\begin{array}{c} 1 \\ 1 \end{array}\right] \frac{e^{-i\frac{Ak}{\nu}\sin\nu\tau}}{\sqrt{2}},\\ &|\Phi^- (\tau)\rangle=e^{-i k^2 \tau}\left[\begin{array}{c} -1 \\ 1 \end{array}\right] \frac{e^{i\frac{Ak}{\nu}\sin\nu\tau}}{\sqrt{2}}, \label{Fstates} \end{aligned}
\end{equation}
that is, the quasienergies $k^{2}$ are doubly degenerate. Using the
basis elements we can construct a determinant similar to that of the
previous section
\begin{equation}
\begin{aligned} &D_2(k)=\\ &\begin{vmatrix} J_m\left(k r_0\right) & N_m\left(k r_0\right) & -J_m\left(k r_0\right) & -N_m\left(k r_0\right)\\ J_{m+1}\left(k r_0\right) & N_{m+1}\left(k r_0\right) & J_{m+1}\left(k r_0\right) & N_{m+1}\left(k r_0\right)\\ J_m\left(k r_1\right) & N_m\left(k r_1\right) & -J_m\left(k r_1\right) & -N_m\left(k r_1\right) \\ J_{m+1}\left(k r_1\right) & N_{m+1}\left(k r_1\right) & J_{m+1}\left(k r_1\right) & N_{m+1}\left(k r_1\right) \end{vmatrix}\end{aligned} \label{tddet}%
\end{equation}
the zero value of which means that boundary conditions given by
Eq.~(\ref{boundary}) are satisfied (see Appendix B). Similarly to the case of constant SOI, the resulting quasienergies $k^2(n,m,\pm)=\epsilon_{n\kappa}^{\pm}$ (and states $|\Phi_{n\kappa
}^{\pm}\rangle$) can be distinguished by specifying the radial ($n$), the angular momentum ($\kappa$) and the spin ($\pm$) quantum numbers. A representative example of the set of the quasienergies is shown
in Fig.~\ref{tdspectrfig}.

Now the problem of the time evolution governed by the Hamiltonian
(\ref{redtdHam}) can be solved for any given initial state $|\psi_{0}\rangle$
using the Floquet basis: One simply needs to expand $|\psi_{0}\rangle$ in this
basis at $\tau=0,$ and use the time dependence of the basis elements to
compute the dynamics.

Let us note that the results above can be adapted to the the case when the SOI strength -- besides the oscillating part discussed above -- has a constant shift, i.e., when $\frac{\omega(\tau)}{\Omega}=B+A\cos(\nu\tau).$ As it is pointed out in the Appendix, we only have to replace $i\frac{Ak}{\nu}\sin\nu\tau$
by $-iB\tau+i\frac{Ak}{\nu}\sin\nu\tau$ everywhere in the exponents.

\section{Time evolution: high harmonics and wave packet motion}
\label{TDsec}
The time dependence of the Floquet states given by Eq.~(\ref{Fstates}) implies that for "allowed" values of $k$ (when the boundary conditions are satisfied) the time evolution of measurable quantities shall contain harmonics of the driving frequency $\nu.$

More explicitly, by using a relevant form of the Jacobi-Anger identity,\cite{AS65} we can write:
\begin{equation}
e^{\pm i \frac{Ak}{\nu}\sin\nu\tau}=\sum_{\alpha=-\infty}^{\infty} J_{\alpha} \left(\frac{Ak}{\nu}\right) e^{\pm i\alpha\nu\tau}.
\end{equation}
This implies, that for SOI oscillations with large enough amplitudes $A,$ several multiples of $\nu$ can appear in the time evolution.

\begin{figure}[tbh]
\includegraphics[width=8cm]{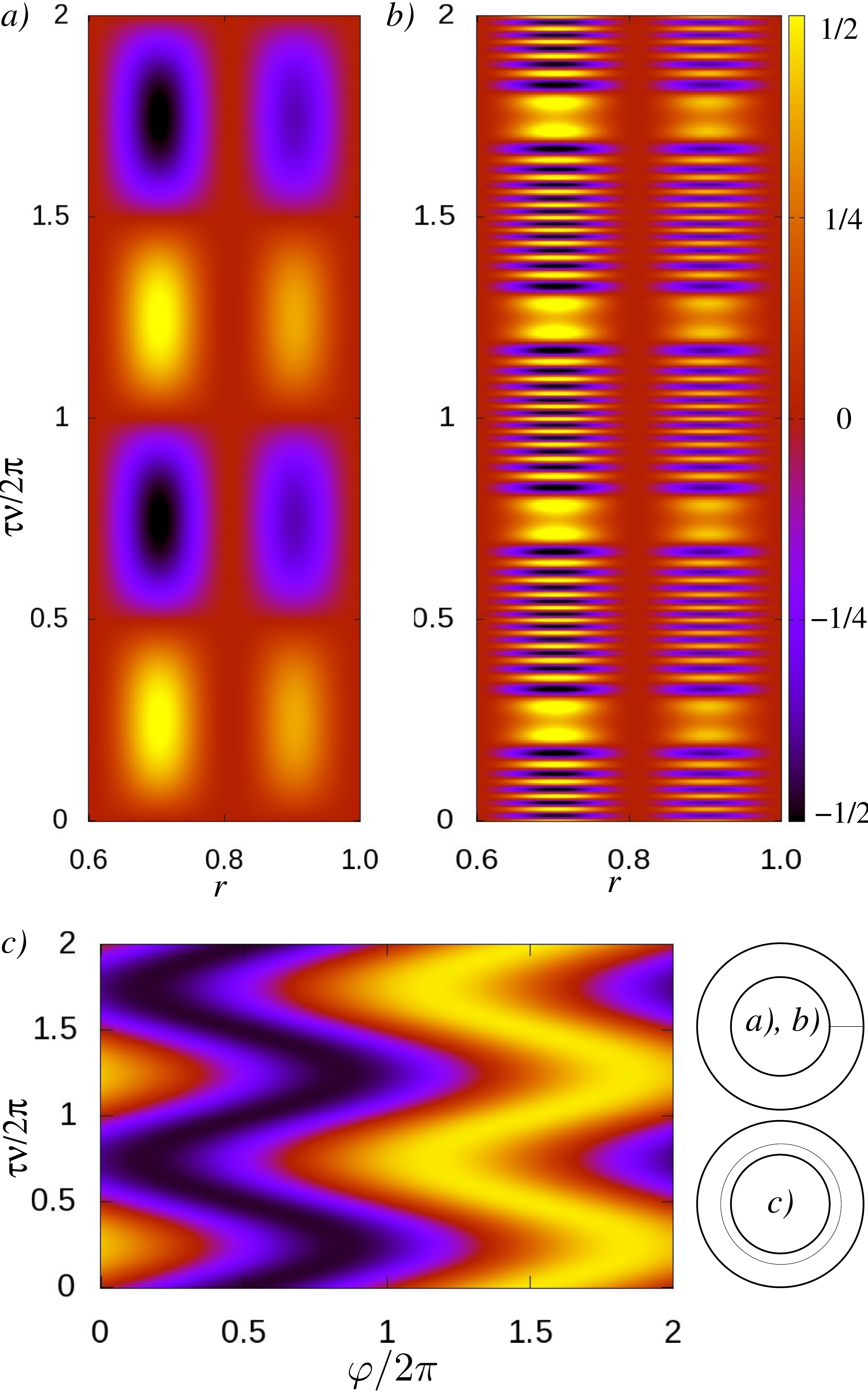}\caption{Spin oscillations in a ring with $r_0/r_1=0.6, \nu=0.01$ as a function of time and coordinates $r$ and $\varphi$. The amplitude of the SOI oscillations is $A=0.1$ for panel (a), where $\langle S_y\rangle(\tau)$ is shown, and $A=3.0$ for panel (b), where we can see the time evolution of $\langle S_x\rangle(\tau).$ The initial state is $\left(\Phi^+_{1,15/2}+\Phi^-_{1,15/2}\right)/\sqrt{2}$ in both cases, and rapid oscillations due to the difference of the quasienergies have been removed by averaging in time domain. Panel c) corresponds to the same parameters and expectation value as a), but it shows a circular section. (See the schematic rings on the right hand side that illustrate what slices are plotted in the different panels.)}
\label{tdspinfig}%
\end{figure}

As an example, we consider doublet states with well-defined $z$ component of the total angular momentum.
The time evolution of these states can be of interest because conservation of the angular momentum may provide a method for preparing them. Absorption of circularly polarized photons e.g.~can excite these states. (Interaction of short
light pulses with the electrons confined in a ring has been discussed in Refs.~[\onlinecite{MB05,ZB08,ZB10}].)
Spin oscillations could be detected e.g. by Faraday rotation experiments.\cite{KMGA05}
Taking a superposition of two states with the same quantum numbers $\kappa$ and $n,$ the resulting spin oscillations
are shown in Fig.~\ref{tdspinfig} (where, in order to focus on the appearance of the harmonics, rapid oscillations resulting from the difference of the quasienergies have been averaged out.) Note the qualitatively different behavior for small and large amplitude SOI oscillations. The amplitude of the spin oscillation shown in Fig.~\ref{tdspinfig} is position dependent, but the frequencies that appear in the time evolution are determined by the amplitude of the SOI variation: Larger amplitude $A$ means that we can see more harmonics of $\nu.$

\begin{figure}[tbh]
\includegraphics[width=8cm]{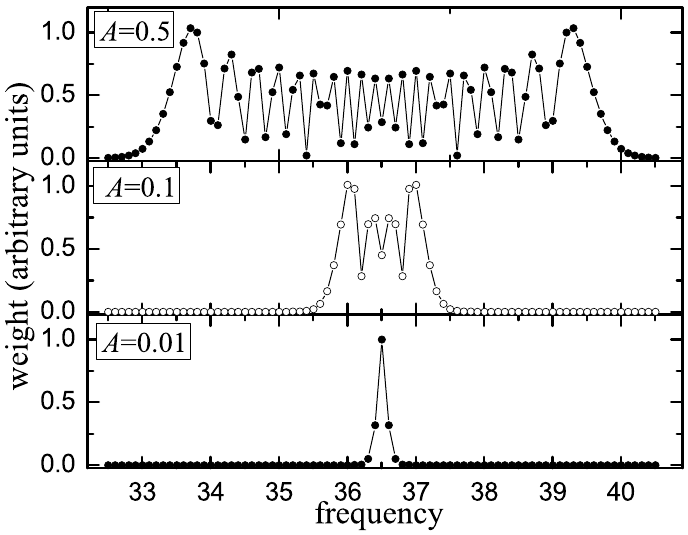}\caption{Normalized magnitudes of the frequency amplitudes in the Fourier series expansion of the probability density corresponding to
$\left(\Phi^+_{1,15/2}+\Phi^+_{2,15/2}\right)/\sqrt{2}$
at $r=0.75, \varphi=0.$} Note that frequency is measured in units $1/\hbar\Omega.$%
\label{fspectrfig}%
\end{figure}

Observable charge density oscillations appear when the initial state is a superposition of spinors corresponding to different radial modes. With realistic parameters, in this case the difference of the quasienergies is larger than routinely achievable driving frequencies. (For a ring with $r_1=250$ nm, $r_0=0.6 r_1$ made of InGaAs, the frequency corresponding to the ground -- first excited radial mode transition is in the THz range.)
Fig.~\ref{fspectrfig} illustrates the broadening of the frequency distribution due the SOI oscillations.  The weight of the various frequency components in the Fourier expansion of
the probability density corresponding to the equal weight superposition of Floquet states with $\kappa=15/2$ and $n=1,2$ is shown in this figure. For small amplitude SOI oscillations, the frequency distribution is narrow, it contains essentially only the difference of the quasienergies, $k_1^2-k_2^2.$ However, larger, but still realistic SOI oscillation amplitudes induce harmonics of the order of a few tens to appear with non-negligible weight, resulting in a considerable widening of the frequency distribution.

\bigskip
Now we consider the time evolution of a localized wave packet where the width of the frequency distribution also plays a prominent role.  Starting with a narrow, spin-polarized initial state, one expects it to spread so that the spin direction changes locally during the process.
However, the boundary conditions imply that when "bouncing back" from the walls, as well as when the "tail" and "head" of the spreading wave packet overlap, we observe interference phenomena. This leads to rather
complex dynamics, which, however, is periodic in the following sense: the discrete nature of the spectrum
(which is a consequence of the boundary conditions and the geometry) may cause the initial phases
to be restored after a certain "revival time". In other words, we expect
periodic "collapse and revival" phenomena: The initially localized wave
packet becomes delocalized along the ring, but later it reassembles again.
\begin{figure}[tbh]
\includegraphics[width=8cm]{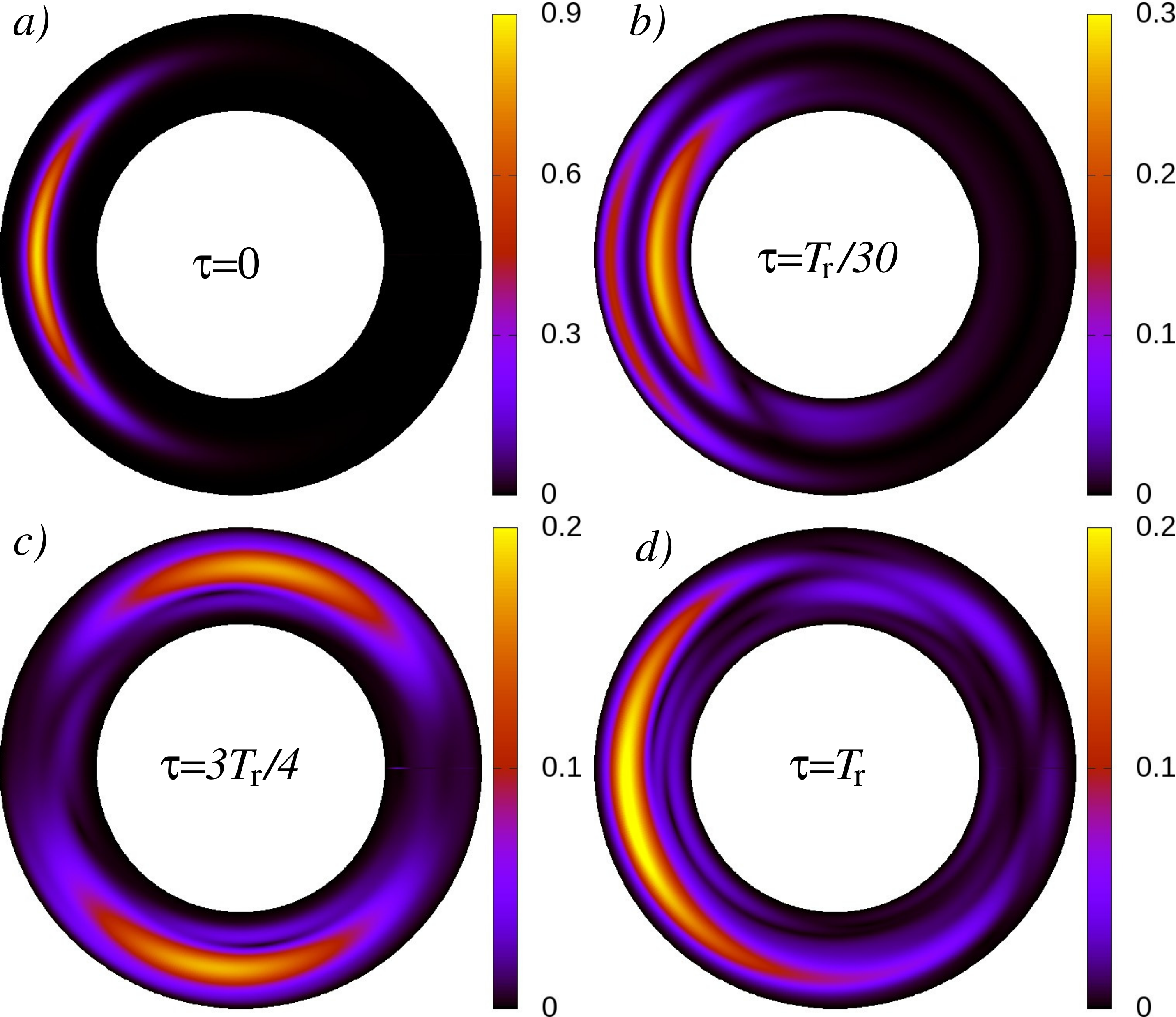}\caption{Collapse and revival phenomenon in a ring with constant SOI ($\omega/\Omega=3.0, r_0/r=0.6.$) The initial state is a Gaussian wave packet, being multiplied by $\sin\frac{\pi (r-r_0)}{r_1-r_0}$ to satisfy the boundary conditions.}%
\label{revfig}%
\end{figure}

Having determined the distribution of the frequencies that appear in a time evolution allows for the estimation of the revival time $T_r.$ For a single-peaked distribution with numerous frequencies involved, the width $\Delta \omega$ of the envelope is the most important quantity, we have $T_r\approx2\pi/\Delta \omega.$ In our case, similarly to the predictions of the one-dimensional model,\cite{FBKP09} the internal structure inside a peak also plays a role leading to a longer revival time as the one that can be deduced from $\Delta \omega.$ However, for initial wave packets narrow enough in the radial direction, several radial modes are excited, leading to rapid phase oscillations. In view of this, we cannot expect perfect revivals, but as the example of Fig.~\ref{revfig} shows, the characteristic features of the collapse and revival phenomena are generally present also in the two-dimensional model. The difference between the dynamics in the radial and azimuthal direction is emphasized by Fig.~\ref{revfig} (b),(c) and (d): superpositions of wave packets localized at different spatial positions ("Schr\"{o}dinger-cat states") appear rapidly in the radial direction compared to the formation of similar states in the azimuthal direction, and approximate revival of the initial wave packet (involving azimuthal coordinates as well) takes also considerably more time. (In fact, Schr\"{o}dinger-cat states in the azimuthal direction appear at integer fractions of $T_r.$) Note that the appearance of "radial Schr\"{o}dinger-cat states" is obviously an interesting two-dimensional feature.

Note that the results of the current paper can be considered as a first step toward the solution of the transport problem through a two-dimensional ring with oscillating SOI strength, which can be of experimental importance. If the leads connecting the ring to the electrodes are narrow enough, we expect the energy levels to be modified only slightly, thus our results can provide a good approximation inside the ring. Considering transport phenomena, besides the usual transversal-mode-related quantization (seen as different parabolas in Figs.~\ref{spectrfig} and \ref{tdspectrfig}) of the conductance, strong SOI oscillations can lead to multiple "sideband currents".\cite{MB03,FBKP09}
That is, qualitatively, the appearance of high harmonics of the driving frequency in the energy spectrum of the transmitted electrons is expected also in the case when wide wires strongly modify the energy spectrum of the ring.
In principle, the spin oscillations coupled to coherent propagation
could be detected also in a closed ring of finite width by an experiment similar to the pump-probe method applied in Ref.~[\onlinecite{KMGA05}].

\section{Summary and conclusions}
\label{Sumsec}
In the current paper we used a wave function picture to describe the dynamics of electrons confined into a two-dimensional ring shaped region, in the presence of oscillating Rashba-type spin-orbit interaction  strength. Our approach is analytic until the point when evaluation of special functions is necessary, it provides the most possible physical insight into the problem. We show that radial boundary conditions are responsible for the discreteness of the Floquet quasienergy spectrum, and determine the corresponding time-dependent eigenspinors. Using this basis it was demonstrated that high harmonics of the SOI oscillation frequency appear in the spectrum of spin and charge density oscillations, already with experimentally achievable SOI oscillation amplitudes. We also determined the time evolution of localized wave packets, collapse and revival phenomena, as well as both radial and azimuthal Schr\"{o}dinger-cat states were shown to appear.

\section*{Acknowledgments}
This work was supported by the Hungarian Scientific Research Fund (OTKA) under Contracts Nos.~T81364, M045596. P.F.~was supported by a J.~Bolyai grant of the Hungarian Academy of Sciences.

\section*{Appendix}

\subsection{Floquet states}
\label{Fapp}
The operator given by Eq.~(\ref{redtdHam}) commutes with itself at any two
time instants $\tau_{1}$ and $\tau_{2}.$ Therefore the time evolution operator
can be written as
\begin{equation}
U(\tau)=U(0,\tau)=e^{-i\int_{0}^{\tau}H(\tau\rq) d \tau\rq}.
\end{equation}
As $H(\tau)=k^2\mathbf{1}+\sigma_x kA \cos(\nu\tau),$ by evaluating the integral we find
\begin{equation}
U(\tau)=e^{-ik^2\tau}\sum_{n} \left(\frac{iAk\sin\nu\tau}{\nu}\right)  ^{n} \frac{{\sigma_{x}}^n}{n!} . \label{U2}%
\end{equation}
Note that in the more general case, when $\frac{\omega(\tau)}{\Omega}=B+A\cos(\nu\tau),$ all the steps described below still can be followed, the single modification is that we have to add $-iB\tau$ to the scalar term in the sum above. (That is, $i\frac{Ak}{\nu}\sin\nu\tau$ should be replaced by $-iB\tau+i\frac{Ak}{\nu}\sin\nu\tau.$)

Since even powers of $\sigma_{x}$ are equal to the $2\times2$ unit matrix,
while the odd ones give $\sigma_{x}$ again, finally we have
\begin{equation}
\begin{aligned} U(\tau)=e^{-i k^2 \tau} \left[\begin{array}{cc} \cos\left(\frac{Ak}{\nu}\sin\nu\tau\right) &-i\sin\left(\frac{Ak}{\nu}\sin\nu\tau\right)\\ -i\sin\left(\frac{Ak}{\nu}\sin\nu\tau\right) &\cos\left(\frac{Ak}{\nu}\sin\nu\tau\right) \end{array}\right]. \end{aligned} \label{U6}%
\end{equation}
The eigenvalues of this matrix are $e^{-i k^{2} \tau\mp i\frac{Ak}{\nu}\sin\nu\tau},$ and
the corresponding eigenvectors $\frac{1}{\sqrt{2}}%
\begin{pmatrix}
\pm 1\\
1
\end{pmatrix}
$ do not depend on time. Comparing the result with Eq.~(\ref{Fbasis}), we
obtain the Floquet states given by Eq.~(\ref{Fstates}).

\subsection{Boundary conditions in the time-dependent case}

$D_2(k)$ given by Eq.~(\ref{tddet}) is obtained by arranging the Floquet states with both possible radial functions (evaluated at the boundaries) in a determinant form. The corresponding linear equation, is, however, time dependent. Its matrix, $M,$ can be obtained from  $D_2$ by multiplying the first two columns  by $\exp(-ik^2\tau-i\frac{Ak}{\nu}\sin\nu\tau)$ while the third and the fourth one by $\exp(-ik^2\tau+i\frac{Ak}{\nu}\sin\nu\tau).$ (The rows of $M$ should in principle also be obtained from that of $D_2$ using the appropriate $\varphi$-dependent factors, but this means only the multiplication of the equations by $\exp(im\varphi)$ or $\exp(i(m+1)\varphi).$) Considering only the signs of the coefficients in the matrices/determinants, we write symbolically
\begin{equation}
D_2(k)=\begin{vmatrix} + & + & - & -\\ + & + & + & +\\ + & + & - & -\\ + & + & + & + \end{vmatrix}.
\end{equation}
Using the same notations and omitting irrelevant common multiplicative factors, the matrix of the homogeneous equation that requires the boundary conditions to be satisfied at any time instant reads
\begin{equation}
M=\cos(\frac{Ak}{\nu}\sin\nu\tau) M_+ - i\sin(\frac{Ak}{\nu}\sin\nu\tau)M_-,
\end{equation}
with
\begin{equation}
M_\pm=\begin{pmatrix} + & + & \mp & \mp\\ + & + & \pm & \pm\\ + & + & \mp & \mp\\ + & + & \pm & \pm \end{pmatrix}.
\end{equation}
The determinant corresponding to the linear equations described by $M_\pm$ is still $D_2,$ thus when it vanishes, both equations have a nontrivial solution vector. Choosing one of them, we have e.g. $M_+ \boldsymbol{\alpha}=\mathbf{0}.$ Obviously, changing the sign of the third and fourth components of $\boldsymbol{\alpha}$, we obtain a solution $\widetilde{\boldsymbol{\alpha}}$ for which $M_-\widetilde{\boldsymbol{\alpha}}=0.$ Finally the spinor valued wave functions satisfying the boundary conditions in the time-dependent case are obtained by using e.g., the first two rows of $M(r)$ ($r_0$ is being replaced by $r$ in $M$):
\begin{equation}
\begin{aligned}
\begin{pmatrix} \psi_1 \\ \psi_2\end{pmatrix}= \cos(\frac{Ak}{\nu}\sin\nu\tau) \sum_n \begin{pmatrix} \left[M_+(r)\right]_{1n}\alpha_n e^{im\varphi} \\ \left[M_+(r)\right]_{2n}\alpha_n e^{i(m+1)\varphi}\end{pmatrix}
\\
- i\sin(\frac{Ak}{\nu}\sin\nu\tau) \sum_n \begin{pmatrix} \left[M_-(r)\right]_{1n}\widetilde{\alpha}_n e^{im\varphi} \\ \left[M_-(r)\right]_{2n}\widetilde{\alpha}_n e^{i(m+1)\varphi}\end{pmatrix}
\\
=e^{-i\frac{Ak}{\nu}\sin\nu\tau} \sum_n \begin{pmatrix} \left[M_+(r)\right]_{1n}\alpha_n e^{im\varphi} \\ \left[M_+(r)\right]_{2n}\alpha_n e^{i(m+1)\varphi}\end{pmatrix}.
\end{aligned}
\end{equation}
According to Floquet's theorem, these states [that correspond to the zeros of $D_2(k)$] multiplied by $e^{-i k^2 \tau}$ can be called the time-dependent eigenstates of the problem.

\end{document}